\documentclass[12pt]{article}
\usepackage{graphicx}
\usepackage[english]{babel}
\usepackage{amsmath}
\usepackage{amsfonts}
\usepackage{amssymb}

\title{Impact of Heater Thermal Properties on Nucleate Pool Boiling: Insights from a Multiscale Automata Simulation}

\author{
Karina I. Mazzitello$^{1}$ $^{2}$,
T. Molina Blanco$^2$,
C. P. Marcel$^1$ $^2$ $^3$,
V. P. Masson$^1$ $^2$
}

\date{}

\begin{document}
\maketitle

\begin{center}
{\small
$^1$ Consejo Nacional de Investigaciones Cient\'{\i}ficas y T\'ecnicas (CONICET) CCT Patagonia
Norte, S. C. de Bariloche, R\'{\i}o Negro (8400), Argentina\\
$^2$ Laboratorio de Termohidr\'aulica, Comisi\'on Nacional de Energ\'{\i}a At\'omica (CNEA),
Bustillo 9500 (8400) S. C. de Bariloche, R\'{\i}o Negro, Argentina\\
$^3$ Instituto Balseiro - Universidad nacional de Cuyo, Bustillo 9500 (8400) S. C. de
Bariloche, R\'{\i}o Negro, Argentina}
\end{center}

\begin{abstract}

This study investigates the influence of heater material properties on nucleate pool boiling using a comprehensive simulation model. Copper and silicon oxide are selected as reference materials due to their properties as excellent and poor heat conductors, respectively. The model integrates well-known heat transfer mechanisms, allowing for the assessment of the effects of these distinct heater materials. The results show that materials with superior thermal diffusivity, such as copper, significantly enhance cooling efficiency during nucleate boiling. Moreover, the study provides insights into the relationship between bubble growth, microlayer recovery beneath a bubble, temperature fluctuations, and heater properties. Comparisons between copper and silicon oxide underscore variations in bubble frequency, attributed to differences in bubble growth time, microlayer recovery time, and material-dependent behavior. The influence of neighboring boiling sites is especially pronounced in silicon oxide due to its low thermal conductivity and diffusivity values. Temperature variations in this material become highly visible due to its very slow response to temperature changes. Simulation results align well with semi-empirical correlations, confirming the model's success in capturing the intricate phenomena of nucleate pool boiling. In summary, the model reveals that changes in the thermal properties of the heater affect not only boiling performance but also key characteristics of the process, including bubble frequency, boiling patterns, regularity, and cavity reactivation speed.

\end{abstract}

\section{Introduction}
\label{Intro}

In contemporary research, power dissipation has emerged as a critical factor across various domains, spanning from the production of miniaturized electronics to nuclear power plants, as highlighted in reviews by \cite{Buongiorno1, Buongiorno2, Buongiorno3}. Within the electronic industry, as MEMS processing technology advances towards high-speed, large buffer memory in small devices, there is a growing focus on the extensive study of pool boiling to achieve efficient cooling. In the nuclear industry, numerous application examples exist. For instance, the design of modern small modular integral reactors (SMiR) also relies on nucleate boiling as a means to achieve self-pressurization \cite{chris13, chris14, chris17b}. \par

Nucleate boiling proves to be effective in dissipating a substantial amount of heat flux with a minimal temperature difference, facilitating efficient component cooling. In pool boiling, heat is transferred from a surface to liquid in a macroscopic state of rest. The initiation of boiling occurs when the local temperature is sufficiently high to permit the formation and growth of vapor bubbles on surface imperfections. These imperfections or cavities with trapped gas act as nucleation sites for the growth of bubbles, arising from vapor initially trapped within them \cite{Buongiorno19}. Surface roughness, porosity, and wettability are observed to influence boiling behavior, although bubble nucleation can also take place on a smooth surface without imperfections. In such cases, nucleation energy is contingent upon the contact angle, defined as the angle between a tangent line to the liquid surface and the solid surface. Typically, a right angle is considered neutrally wetting, with lesser angles indicating hydrophilic properties and greater angles indicating hydrophobic characteristics. Both hydrophilic and hydrophobic surfaces exhibit some positive contributions to improving boiling performance, with hydrophilic surfaces aiding in rewetting following bubble departure, and hydrophobic surfaces promoting bubble nucleation seeded by vapor stored in cavities \cite{Buongiorno1, Buongiorno2}. Additionally, the contact angle between the liquid-vapor interface at the bubble base and the surface varies during the stages of bubble growth and departure affecting both the size of the bubbles and the departure frequency \cite{Wang19,18Wang19, 30Wang19, chris17}.\par

Recent experimental research in nucleate pool boiling has emphasized the development of novel surface manipulation methods to enhance phase-change heat transfer. These methods include mechanical machining, chemical treatments, nanoparticle coatings, and micro-/nanoelectromechanical systems techniques, such as photolithography and reactive ion etching, as well as fiber-laser texturing that avoids additional layers \cite{20Golorcic18, 21Golorcic18, 22Golorcic18, 23Golorcic18, 24Golorcic18}. \par

While numerous numerical models have been developed in an attempt to explain experiments exploring the bubble cycle \cite{4cho19, 5cho19, 6cho19}, predicting boiling heat transfer remains intricate due to the necessity of considering phenomena occurring over multiple scales, from the adsorbed vapor at the nanometer scale \cite{4guion18, 5guion18} to the bubble diameter at the millimeter scale. Timescales also vary widely, from microlayer formation at the microsecond scale ($10-100\; \mu s$) to evaporation at the millisecond scale ($1-10\; ms$), significantly contributing to bubble growth \cite{18Guion18, 182guion18,16guion18,162guion18,15guion18}.

In the past, several researchers attempted to model the pool boiling curve numerically, employing semi-analytical methods and continuum techniques like volume-of-fluid and level-set methods which were reviewed in \cite{Dhir13}. Fogliatto et al. \cite{Fogliatt21} used the lattice Boltzmann method, revealing trends in nucleation temperature concerning heat flux and contact angle. While continuum numerical approaches such as volume‑of‑fluid and level‑set methods have significantly advanced our understanding of nucleate and film boiling, they remain computationally expensive and often struggle to predict boiling on surfaces with complex patterns or non-uniform wettability \cite{Abar04, Malan15, Trygg09}. Direct Numerical Simulation (DNS) offers the highest level of fidelity, resolving microlayer formation, bubble dynamics, and conjugate heat transfer by solving the full Navier-Stokes and energy equations without empirical closures. For example, Bure{\v{s}} and Sato \cite{Sato22} explicitly resolved microlayer dynamics and heat conduction between fluid and solid domains, demonstrating quantitative agreement with experiments albeit at very high computational cost. A more recent study by Long et al. \cite{Long25} employed adaptive mesh refinement (AMR) in Basilisk, an open-source CFD solver designed for multiphase flow simulations on adaptive grids, to simulate a full bubble cycle with resolved microlayer and conjugate heat transfer, reducing cost by three orders of magnitude compared to earlier DNS efforts, yet still requiring significant resource investment. Despite these advances, full-scale DNS remains infeasible for engineering domains. By contrast, our automata‑based model bridges the gap between semi‑empirical boiling correlations and full CFD/DNS approaches by capturing spatial interactions among nucleation cavities and surface thermal response at drastically reduced computational cost. In this context, the term automata refers to a coarse-grained computational framework in which discrete surface cells interact through physically motivated local rules governing heat transfer and bubble dynamics. The collective dynamics of bubble nucleation, growth, and interaction place this system within the broader context of soft condensed matter, where emergent behavior arises from many-body interactions far from equilibrium.

Numerical simulations using 2D cellular automata \cite{Jing03} and 3D coupled lattice \cite{Gupta06} showed success but struggled with bubble-population dynamics in rigid-grid representation. Herrero et al. \cite{Herrero96} proposed geometric automata, where bubbles are simulated as interacting disks. Later, Marcel et al. \cite{Marcel11} successfully extended this paradigm to 3D models of automata for pool boiling. Furthermore, this model was employed to simulate experimental boiling data obtained from hydrophilic and hydrophobic heating surfaces, demonstrating excellent agreement \cite{chris17}.

In the context of ongoing efforts to improve heat transfer, the present study investigates the impact of thermophysical properties on nucleate boiling for two materials with distinct characteristics: copper and silicon oxide. Notably, copper exhibits thermal diffusivity and conductivity two orders of magnitude greater than silicon oxide ($\alpha _{Cu} = 116.5967\; mm^2/s$ vs. $\alpha_{SiO_2} = 0.8344\; mm^2/s$; and $\kappa_{Cu}= 401\;W/mK$ vs. $\kappa_{SiO_2}= 1.38\;W/mK$ respectively). For such a task, the model developed in this work is based on \cite{Marcel11} and \cite{chris17} while it incorporates some refinements intended to capture more accurately the dynamics of both the bubble growth and the superheated microlayer. Moreover, the model is highly versatile, spanning multiple scales from nano to milli, and timescales from tenths of milliseconds to minutes, allowing the independent variation of parameters through computational simulations.

Experimentally reproducing boiling data is a challenge due to contamination and cavity reactivation. In this work, we explore - through an extended automata model - how copper and silicon oxide substrates affect boiling frequency, temperature regularity, and cavity activation behavior, while maintaining fully controlled cavity distributions and fixed contact angle. The results provide both mechanistic insight and design-oriented performance metrics for materials under nucleate boiling conditions.

\section{Model}
\label{Model}
\begin{figure}[htp]
\centering
\includegraphics[width=20pc]{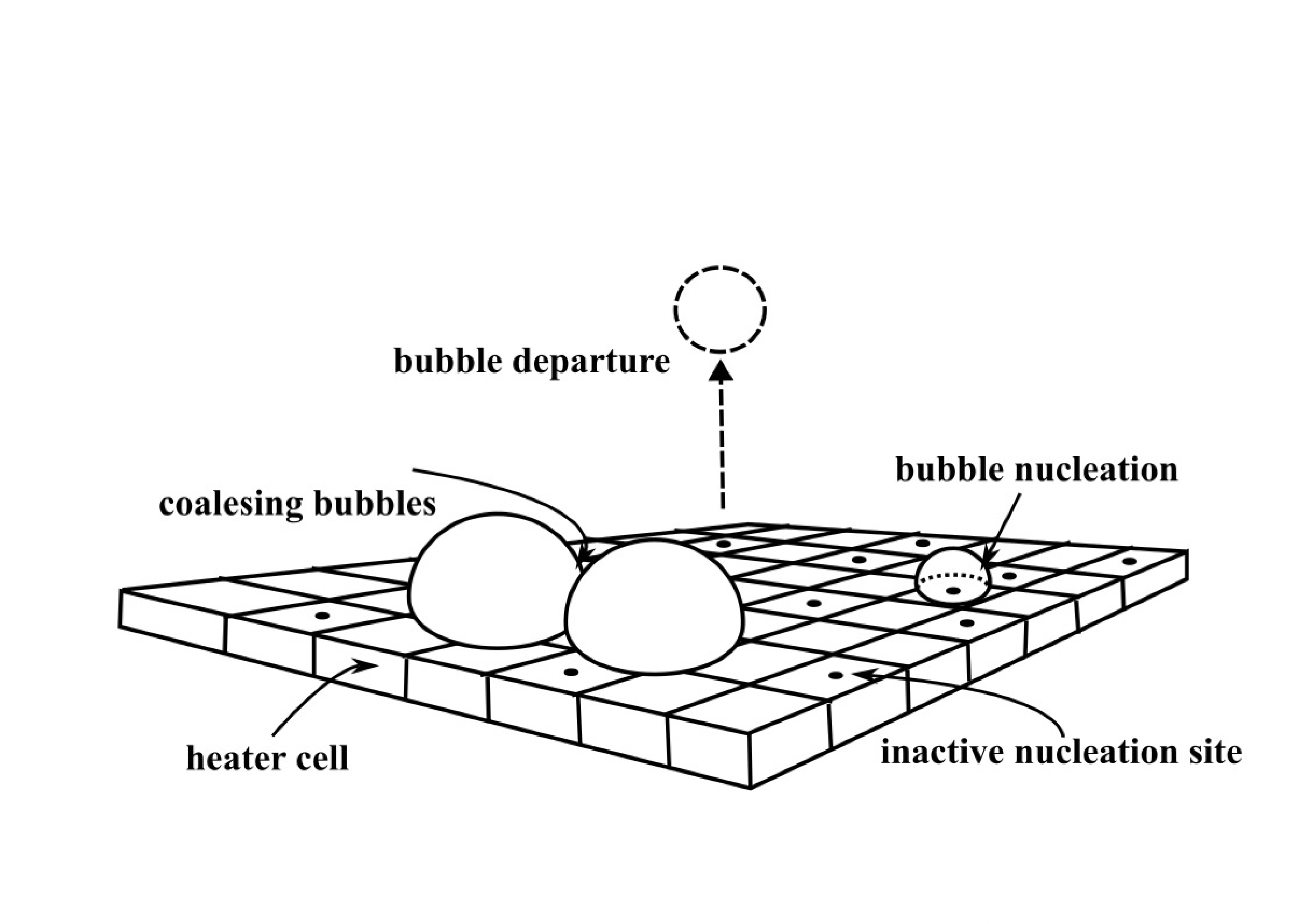}  
\caption{\label{Esquema} Model setup diagram. Black dots represent cavities, while hemispheres and circles depict bubbles. 
}
\end{figure}

Consider a scenario of pool boiling, where a small heater plate is positioned horizontally in a container filled with stagnant saturated water at atmospheric pressure. The model consists of two distinct domains operating at different scales: the heat conduction process within the heater and the upper boiling two-phase flow including the growth and departure of bubbles from the surface (see Fig. \ref{Esquema}). These domains are coupled through the heat transfer mechanisms involved in surface cooling. The square heater, measuring $1\times 1 $ $cm^2$ with a thickness of $0.25\;cm$, discretized into cubic cells of $250\times 250 \times 250\;\mu m^3$ (refer to Eq. (\ref{Fourier})), resulting in a total of $16,000$ cells. It is worth noting that simulations with larger heaters, up to $5\times 5\;cm^2$ in area and $1\;cm$ in thickness, yielded results identical to those presented in this study.

Cavities with a size following Gaussian distributions with a mean of $6 \pm 3 \; \mu m$ are randomly placed at centers of $20$ randomly selected cells (cavities with a radius less than zero are excluded). A subset of them may become activated based on Eq. (\ref{rminrmax2}), allowing vapor bubbles to develop. A fixed contact angle of  $\phi = 20^\circ$ between the water-vapor interface and the heater surface is adopted. Bubbles grow and depart or can coalesce with other bubbles on the surface before departure. Since the investigated pool nucleate boiling regime is far from its critical state, the model does not consider the subsequent behavior of bubbles after departure.

\subsection{Heat conduction in the heater}
\label{Heat_conduction}

Starting with the Fourier equation given by

\begin{equation}\label{Fourier}
\frac{1}{\alpha_m}\frac{dT}{d\tau} = \overline{\nabla}^2 T + \frac{\dot{q}}{\kappa_m}  \;\;\mbox{,} 
\end{equation}
the temperature field in the heater is estimated using the alternating-direction implicit (ADI) finite difference method, which is modified by an $f$ factor ($0<f<1$) as proposed in \cite{Chang91}. In Eq. (\ref{Fourier}), $\alpha_m$ and $\kappa_m$ represent the heater's thermal diffusivity and conductivity, respectively, and $\dot{q}$ denotes the volumetric heat sources and/or sinks, per unit of time.\par

In each grid cell of the heater, the heat fluxes due to the first term in Eq. (\ref{Fourier}) are affected by the factor $f$, remaining the total heat flux according to Eq. (\ref{Fourier}) in each direction over a complete time step. The advantage of the modified ADI method lies not only in the need to solve only tridiagonal matrices, but also in allowing significantly larger time steps without compromising convergence or stability of the solution. The time step limit for the conventional ADI method can be increased by a factor of $1/f$ through the use of this modified ADI method. For example, the time step limit can be increased by one order of magnitude with $f = 0.1$, and the solutions still remain stable with high accuracy (see \cite{Chang91} for more details).\par

The heat generated within the heater is extracted from its top by the liquid. The extraction mechanisms include radiation, microconvection, microlayer evaporation due to bubble growth, and natural convection generated by the fraction of the wetted cell surface in contact with the liquid without bubbles. Although its contribution sometimes is negligible radiation is always present, whereas the other extraction mechanisms are mutually exclusive. In essence, microconvection and microlayer evaporation are relevant processes for heat extraction despite intermittently assisting natural convection (refer to Section \ref{Heat_transfer}). As these mechanisms undergo relatively rapid changes over time, solving Eq. (\ref{Fourier}) using the ADI method is advantageous, especially since the precision is on the order of $2$ in time ($O(t^2)$), allowing for better tracking of temporal changes compared to explicit finite difference methods. We employ time steps of $10^{-3}$ to $10^{-4}$ seconds, depending on the case, to account for rapid processes.

\subsection{Heat transfer mechanisms}
\label{Heat_transfer}

\begin{figure}[htp]
\centering
\includegraphics[width=18pc]{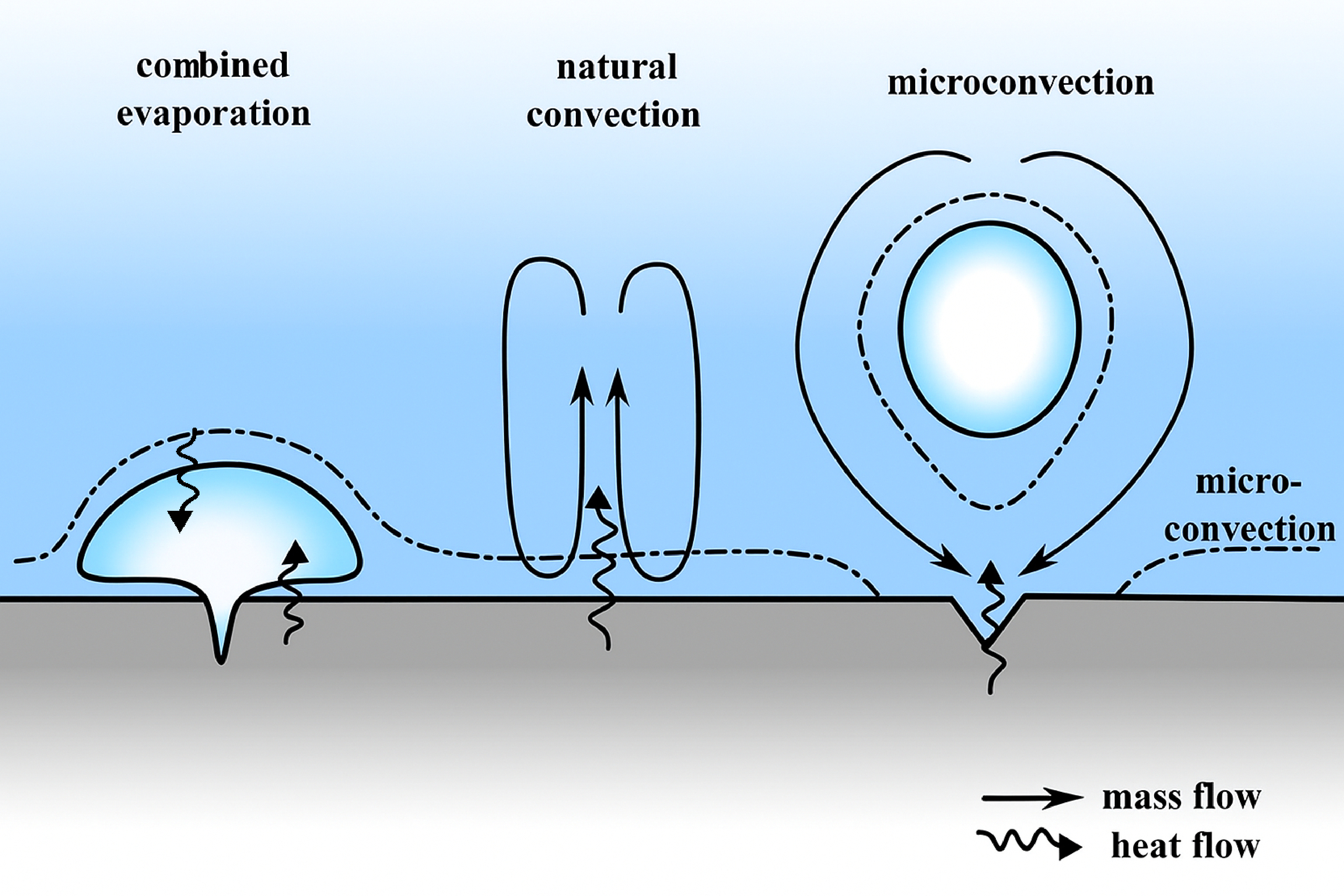}  
\caption{\label{mechanisms} Schematic representation of the boiling heat transfer model, including {combined microlayer and superheated liquid evaporation}. Radiation heat is not depicted due to its minor contribution in nucleate boiling.}
\end{figure}

The mechanisms responsible for extracting thermal energy from the heater and transferring it to the liquid are illustrated in Fig.\ref{mechanisms}:

\begin{itemize}
    
    \item Natural convection ($q_{nc}$): The density difference between the hot and cold liquid induces convective flows,
    extracting heat from the heater across its free area, $A_{nc}$.
    
    \item Combined microlayer and superheated liquid evaporation ($q_{me}$): It is assumed that the vapor mass forming the bubble originates from both the evaporation of the superheated microlayer between the heater surface and the bubble, and from the evaporation of the superheated liquid surrounding the bubble periphery.
    
    \item Microconvection, ($q_{mc}$): Detaching bubbles remove the superheated liquid layer in their vicinity. Colder liquid from the bulk of the pool quenches an area $A_{mc}$, and heat is required to recover the microlayer 
    superheating.
    
    \item Radiation, ($q_{rad}$): Heat is transferred from the heater to the liquid via electromagnetic radiation.
\end{itemize}

The total heat rate extracted from an upper cell to the liquid, $\dot{q}_{out}$, is given by the sum of individual contributions:

\begin{equation}\label{q_out}
\dot{q}_{out} = \dot{q}_{nc} + \dot{q}_{mc} + \dot{q}_{me} + \dot{q}_{rad} \mbox{.} 
\end{equation}
The net heat rate extracted from top cells represents the difference between the power generated in the cell, $\dot{q}$ (see Eq. (\ref{Fourier})), the heat rate transferred to or from adjacent cells through conduction and the extracted heat rate by the mechanisms described in Eq. (\ref{q_out}). If this net heat rate is different from zero, the temperature in the cell will vary. The extracted heat from the inner cells through mechanisms shown in Eq. (\ref{q_out}) is zero since they are not wetted by the fluid.\par

The radii of nucleation sites (cavities) are randomly selected and placed at the centers of 20 top cells of the heater. The cavity sizes are sampled from a normal distribution with a mean of $6;\mu m$ and a standard deviation of $3;\mu m$ \cite{chris17}. At the active sites, spherical cap bubbles grow attached to the cells until reaching a critical detachment radius, $r_d$. At this point, spherical bubbles depart (see Fig~\ref{mechanisms}), initiating the growth process again once the site becomes reactivated. \par 

If a growing bubble touches other bubbles on the surface, they coalesce to form a single bubble at the nucleation site of the largest one, preserving the amount of vapor (see Fig. \ref{Esquema}).\par

\subsubsection{Effect of the contact angle and natural convection}
\label{angle}

The current model incorporates the influence of the contact angle $\phi$ on the site activation process through the activation rule. Accordingly, a given site is active if its radius $r$ satisfies \cite{VanStralen75}:

\begin{equation}\label{rminman}
 r_{min} < r < r_{max}\;\mbox{,}
\end{equation}
where
\begin{equation}\label{rminrmax2}
 r_{min,max} = \frac{\delta}{2C_1}\left\{ 1 - \frac{\theta_s}{\theta_w} \pm 
 \sqrt{\left[ 1 - \frac{\theta_s}{\theta_w}\right ] ^2 - 
 \frac{4\zeta C_3}{\delta \theta_w}}\right\}
\end{equation}
with

\begin{equation}
\zeta = \frac{2\sigma T_{sat}}{\rho_v h_{lv}},\;\;\;\;
C_1 = \frac{1 + cos \phi }{sin \phi } \;\;\mbox{and} \;\;C_3 = 1 + cos \phi \;\mbox{,}
\end{equation}

where\par
$\delta$: thermal boundary layer thickness\par
$\phi$: contact angle between a tangent to the liquid
surface and the solid surface\par
$\theta_s$: liquid subcooling, i.e. $T_{sat}-T_{\infty}$ (equal to zero)\par
$\theta_w$: wall superheat of the cell containing the nucleation site, i.e. $T_w-T_{sat}$\par
$\rho_v$: vapor density\par
$h_{lv}$: latent heat of vaporization\par
$T_{sat}$: saturation temperature\par
$T_{\infty}$: liquid bulk temperature\par
$\sigma$: surface tension.\par

The thermal boundary layer thickness is estimated as $\delta=\kappa /\overline{h}$, where $\kappa$ is the liquid heat conduction coefficient, and $\overline{h}$ is the average heat transfer coefficient assessed using the average temperature of the heater. The average heat transfer coefficient is given by \cite{mcadams54,goldstein73,lloyd74}:

\begin{equation}\label{h_conv}
\overline{h}= \frac{\kappa}{L} \overline{Nu} \;\;\mbox{,}
\end{equation}

where $L$ is the heater characteristic length, $L = 4\frac{A_L}{P}$, with $A_L$ and $P$ being the heater upper area and perimeter, respectively. $\overline{Nu}$ is the free convection Nusselt number calculated as:

\begin{equation}\label{Nu}
\overline{Nu} = \left\{ \begin{tabular}{ l l}
              $0.54 c_{nc} Ra^{1/4}$, & \mbox{if} $10^4 \le Ra_L < 10^7$ \\
              $0.15 c_{nc} Ra^{1/3}$, & \mbox{if} $10^7 \le Ra_L < 10^{11}$ \\
             \end{tabular} \right.
\end{equation}

where the Rayleigh number is defined as:

\begin{equation}\label{Ra}
 Ra = \frac{g \beta \left( T_w - T_{\infty} \right) L^3}{\nu \alpha_l} \;\mbox{,}
\end{equation}

with $g$ being the gravitational constant, $\beta$ the thermal expansion coefficient of the liquid, $\nu$ the kinematic viscosity, and $\alpha_l$ the thermal diffusivity of the liquid. The coefficient $c_{nc}$ is included to account for border effects in small heaters and extra agitation due to the detachment of nearby bubbles. In the present model, $c_{nc}$ is taken to be equal to 0.4, as suggested in \cite{chris17}.\par

Finally, natural convection heat transfer acts on the portion of the heater surface in direct contact with the liquid, as shown in Fig. \ref{mechanisms}. For every top cell of the heater, the convection heat transfer rate is given by:

\begin{equation}\label{q_nc}
\frac{dq_{nc}}{dt} = c_{nc} \overline{h} A_i \left( T_w - T_{\infty} \right)\;\mbox{,} 
\end{equation}
where $A_i$ is the area of cell $i$ in contact with the liquid.\par

\subsubsection{Microlayer Evaporation, Superheated Liquid Evaporation, and Microconvection}

The vapor mass that forms the bubble originates from two primary sources: (1) evaporation of the superheated microlayer between the heater surface and the bubble, and (2) evaporation of the superheated liquid surrounding the bubble \cite{Yabuki14}. This process occurs shortly after the bubble reaches its equilibrium radius. Heat is transferred through both the microlayer and the superheated liquid, and their thicknesses progressively decrease due to evaporation, which can be described by:

\begin{equation}\label{q_me}
 q_{me} = \frac{4}{3}\pi \left( r(t+\Delta t)^3 - r(t)^3 \right) \rho_v h_{lv} \;\mbox{,}
\end{equation}
where $r(t)$ is the bubble radius at time $t$, and $\Delta t$ is the time step. The temporal evolution of the bubble radius is modeled following Van Stralen et al. \cite{VanStralen75}:
\begin{equation}\label{r}
 r(t) = \frac{1}{1/R_1 + 1/R_2}
\end{equation}
In this expression, $R_1$ dominates during the initial growth stage and is given by:
\begin{equation}\label{R1}
 R_1 = 0.8165 \sqrt{\frac{\rho_v h_{lv}}{\rho_l T_{sat}} \left(T_{cavity} - T_{sat}\right ) exp\left(-\sqrt{t/t^*_g}\right)} \; t 
\end{equation}
At later times, $R_2$ becomes the dominant contribution, describing the effect of microlayer and superheated liquid evaporation:
\begin{equation}\label{R2}
R_2 = \sqrt{\frac{12}{\pi}} \left(b^*f_1 + 0.373f_2Pr^{1/6}\right) Ja \sqrt{\alpha_l t} exp\left(-\sqrt{t/t^*_g}\right)\;\mbox{.}
\end{equation}
In these equations:
\begin{itemize}
 \item $R_1$ (Eq. \ref{R1}) represents the contribution from liquid inertia.
 
 \item $R_2$ (Eq. \ref{R2}) accounts for combined microlayer evaporation, where:
 
 \begin{itemize}

    \item The first term (weighted by $f_1$) corresponds to evaporation from the microlayers surrounding the bubble.

    \item The second term (weighted by $f_2$) corresponds to evaporation from the microlayer beneath the bubble.
 \end{itemize}
\end{itemize}

The parameter $t^*_g$ denotes the bubble growth time; $b^* $ is the fraction of the bubble's surface area in contact with superheated liquid; $Pr$ is the Prandtl number; and $Ja$ is the Jakob number. In our simulations, we used fixed values of $b^* =0.2$, $f_1=0.6$, and $f_2=0.4$, while $t^*_g$ is dynamically computed during simulation.

Equation (\ref{r}) remains valid until the bubble reaches a critical departure radius $r_d$, determined by the Fritz stability criterion \cite{Fritz35}:

\begin{equation}\label{r_d}
r_d = \frac{1}{2} 0.0148 \phi \sqrt{\frac{2\sigma }{g\left( \rho_l - \rho_v \right)}} \;\mbox{,} 
\end{equation}

When a bubble detaches from the heated surface, it removes part of the superheated thermal layer. Subsequently, colder liquid from the bulk quenches the surface. If the nucleation site remains active, a new superheated thermal layer is formed (see Fig. \ref{mechanisms}). The heat removed through microconvection due to bubble departure is modeled as \cite{Lahey}:

\begin{equation}\label{micro}
 q_{mc} = c_l\rho_l \frac{2}{3} \pi r_d \left( c_{mc} r_d \right)^2
 \left( \frac{T_w - T_{\infty }}{2}-T_{\infty }\right) \;\mbox{,}
\end{equation}
where $c_l$ is the specific heat of the liquid, and $c_{mc}$ is the microconvection coefficient representing the volume of liquid entrained with the departing bubble. Lahey et al. \cite{Lahey} suggested a value of $c_{mc}=1.3$; however, in our model we set $c_{mc}=1.2$, as this choice yields better agreement with recent experimental data \cite{chris17}.

The waiting time, $t_w$, is defined as the period from the onset of thermal layer formation until the beginning of bubble growth. Han and Griffith proposed a minimum waiting time based on the critical thickness of the thermal layer required for bubble nucleation. This minimum value is estimated as:

\begin{equation}\label{twmin}
(t_w)_{min} = \frac{144{\left(T_{cavity}-T_{\infty}\right)^2}T_{sat}^2 \sigma^2}{\pi\alpha_l\rho_v^2h_{lv}\left(T_{cavity}-T_{sat}\right)^4}\;\mbox{,}
\end{equation}
where $T_{cavity}$ is the cavity temperature at the moment of bubble departure. In our simulations, we assume that the actual waiting time is at least on the order of $t_g$, and we adopt a effective value of $t_w=20 (t_w)_{min}$. During this recovery time, no bubble growth occurs, and the cavity remains inactive, while heat continues to be removed via microconvection ($q_{mc}$).

\subsubsection{Radiation}
\label{Radiation}
The power removed through the radiation process is accounted for in the model by

\begin{equation}
 \frac{dq_{rad}}{dt} = A_T \epsilon \sigma_{sb} \left( T_w^4 - T_{\infty}^4 \right) \;\mbox{,}
\end{equation}
where $A_T$ is the total area of the cell, $\epsilon$ is the emissivity of the upper surface of the heater (assumed to be equal to 0.8 in the present model), and $\sigma_{sb}$ is the Stefan-Boltzmann constant ($\sigma_{sb}=5.67\times 10^{-8}[W/m^2K^4]$)

\section{Simulation results}
The simulation begins with the silicon dioxide and copper heaters initially in thermal equilibrium with the saturated water and then start producing power for 60 seconds. Then the simulation stops. The system thus exhibits two macroscopic stages: a transient stage and a so-called steady-state stage. The steady state condition will be defined as the state in which macroscopic statistical parameters remain unchanged, as it will be done explicitly further in
this section.\par
 It is important to note that these results are applicable under the assumption of a fixed contact angle and constant cavity densities with equivalent Gaussian size distributions. In practical experiments, maintaining constant surface properties can be challenging. In essence, the model facilitates the systematic study of this phenomenon by varying one parameter at a time. \par
 
\subsection{Temporal evolution: from thermal equilibrium to boiling steady-state}

\begin{figure}[htp]
\centering
\includegraphics[width=17pc]{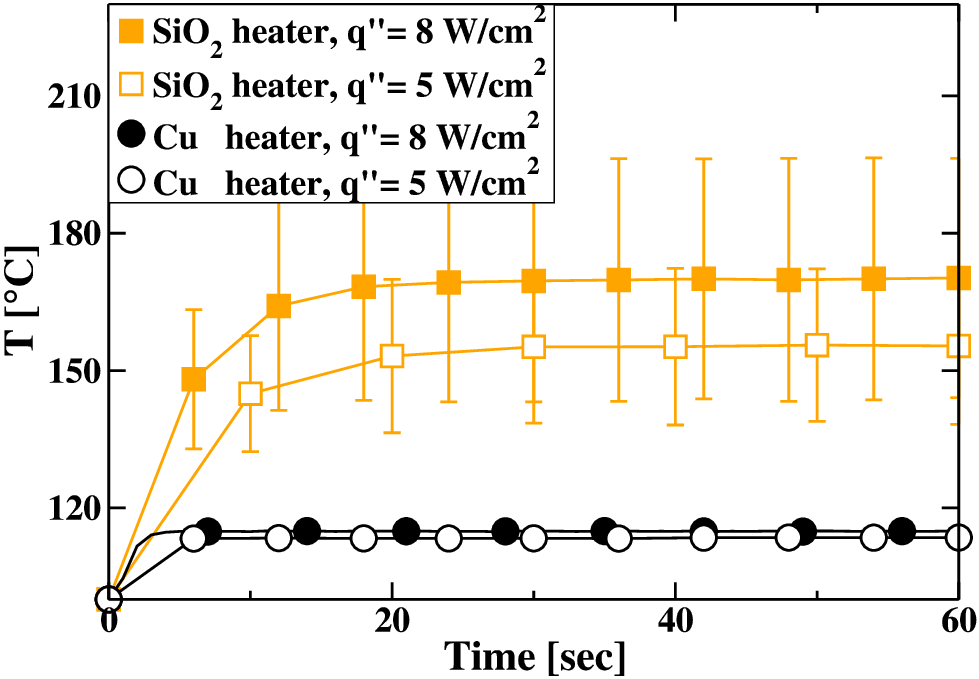}
\caption{\label{Temp_Pot5_8} Temporal evolution of the average surface temperature found in the $SiO_2$ and the $Cu$ heaters. For reference, error bars have been plotted, indicating the standard deviation of the average surface temperature across all surface cells in the heaters.}
\end{figure}

The complete temporal evolution of the average surface temperature for the $SiO_2$ and $Cu$ heaters was computed for applied power fluxes of $5$ and $8\;W/cm^2$, and the results are presented in Fig.~\ref{Temp_Pot5_8}. Error bars denote the standard deviation of the average surface temperature across all surface cells of each heater.

A comparison of the heaters' performance reveals that the $Cu$ heater exhibits a significantly shorter transient phase than the $SiO_2$ heater. This faster thermal response can be attributed to the substantially higher thermal diffusivity of copper, approximately two orders of magnitude greater than that of $SiO_2$. As a result, under steady-state conditions, the $Cu$ heater maintains a relatively low and spatially uniform surface temperature. In contrast, the $SiO_2$ heater reaches higher average surface temperatures with increased spatial variability, influenced by bubble departure dynamics and its slower thermal response, as further discussed in this section.

\begin{figure}[htp]
\centering
\includegraphics[width=17pc]{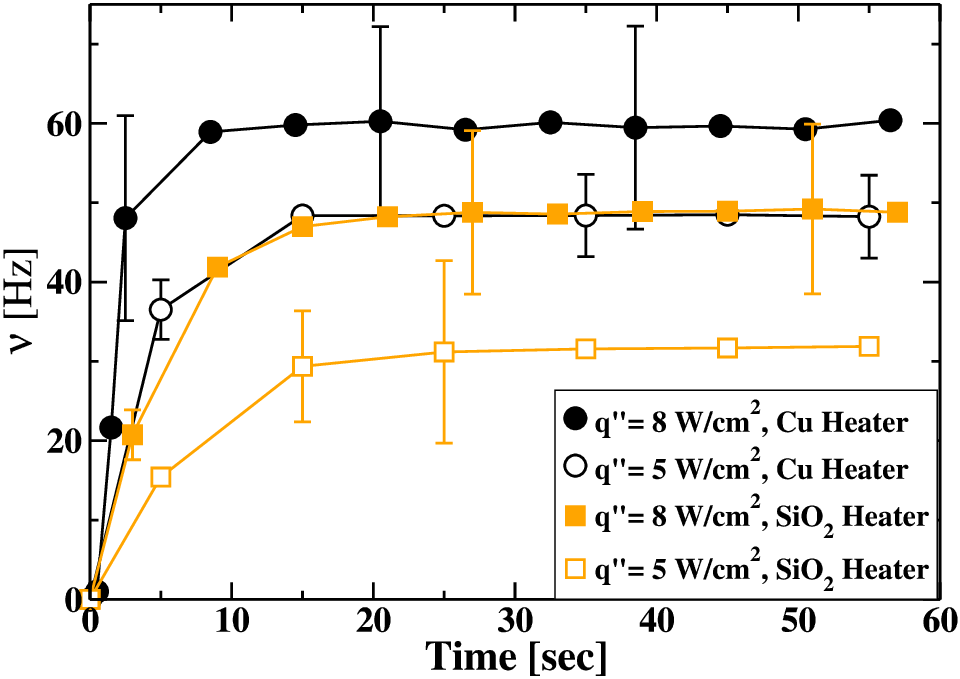}  
\caption{\label{Freq_media} Mean boiling frequency evolution over time up to reach the steady state for two different powers, for both $Cu$ and $SiO_2$ heaters. The error bars in the plot indicate the standard deviation of the average boiling frequency across all active cells in the heaters. }
\end{figure}

\begin{figure}[htp]
\centering
\includegraphics[width=17pc]{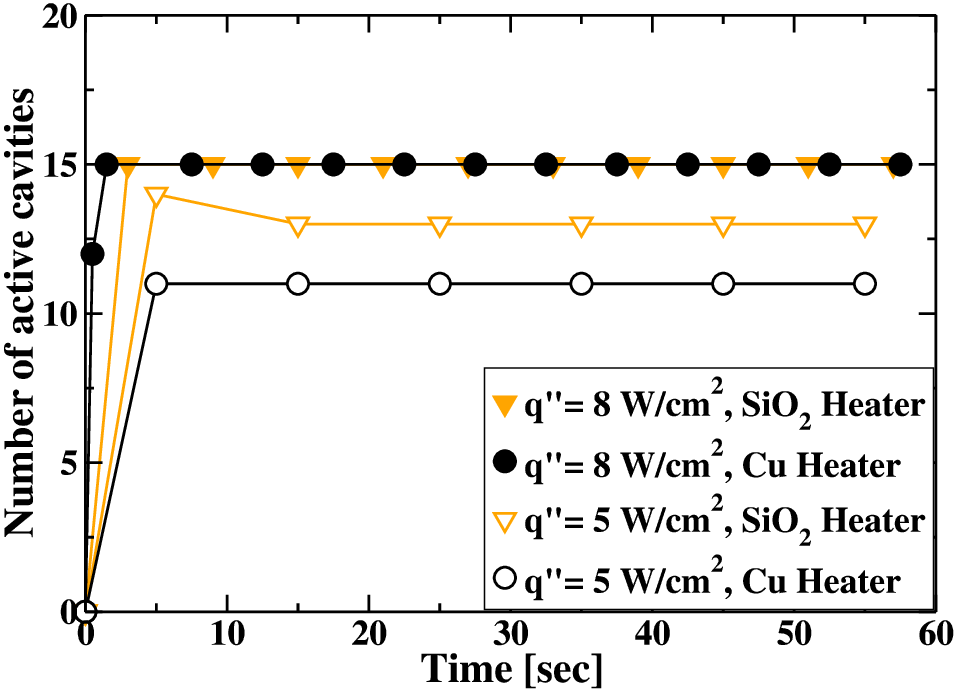}  
\caption{\label{number_active_sites} Evolution the number of active cavities vs. time up to reach the steady state for two different powers, for both $Cu$ and $SiO_2$ heaters.  }
\end{figure}

To further investigate the boiling dynamics and associated topological behavior, we analyzed the time evolution of two quantities: (i) the number of active cavities producing vapor, and (ii) the mean bubble departure frequency. These are shown in Figs. \ref{Freq_media} and \ref{number_active_sites} for applied power fluxes of $5$ and $8\;W/cm^2$.

The mean frequency was computed using a moving average over time windows ranging from $6$ to $10$ seconds, 
depending on the timescale of the events. The boiling frequency is influenced by both the bubble growth time 
and the delay between successive bubble generations at each cavity, which in turn are affected by heat transfer 
processes within the heater material. 

As seen in Fig. \ref{Freq_media}, the transient period is again shorter for the Cu heater, reflecting its faster thermal response to the applied power flux. This difference is especially pronounced at $q''=5\;W/cm^2$. Once steady-state is reached, the $Cu$ heater consistently shows higher mean departure frequencies than the $SiO_2$ heater for any given power flux. This is likely due to the slower re-establishment of the superheated microlayer on the $SiO_2$ heater after bubble detachment, which delays the reactivation of the nucleation site.

Conversely, the behavior is reversed when analyzing the number of active nucleation sites. At $5\;W/cm^2$, the $SiO_2$  heater shows a higher number of active cavities compared to the $Cu$ heater, while at $8\;W/cm^2$ the number becomes approximately equal (see Fig. \ref{number_active_sites}). This trend aligns with the observation that the surface temperature of the $SiO_2$ heater remains significantly higher than that of the $Cu$ heater throughout the simulation, thus activating a larger number of nucleation sites.

\begin{figure}[htbp]
\centering
\includegraphics[width=34pc]{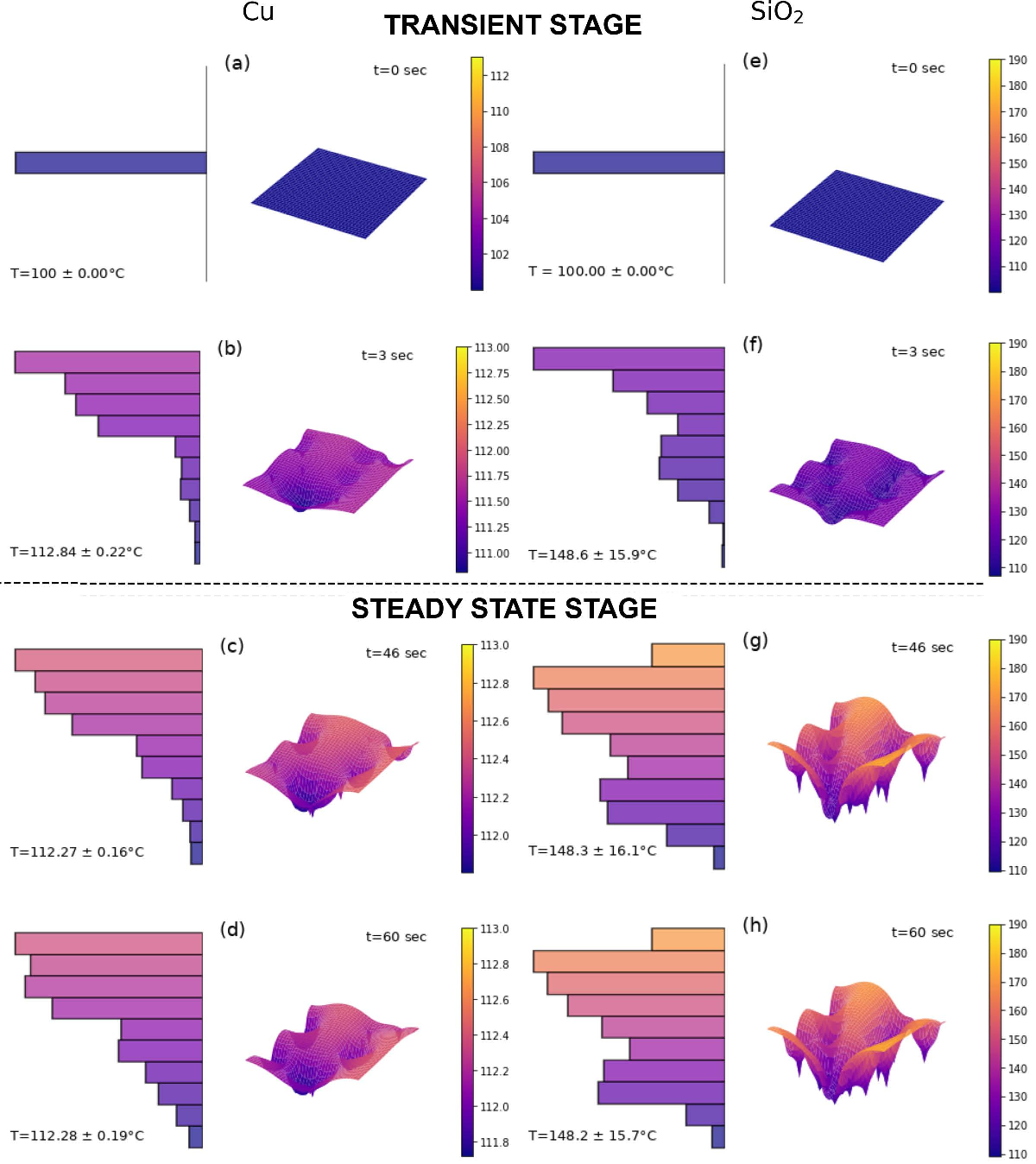}  
\caption{\label{Temp-Bubbles} Temporal evolution of the surface temperature and its normalized histogram for a $Cu$ heater (a)-(d) and a $SiO_2$ heater (e)-(h), both initially at the saturation temperature of water when applying a power flux equal to $q''=5\;W/cm^2$ ((a) and (e)).  The colorbar on the right side of the plots provides the temperature scale.}
\end{figure}

Fig. \ref{Temp-Bubbles} shows the temporal evolution of the surface temperature and its histogram for both materials at different times, covering the transient, parts (a) and (b) for $Cu$ and (e) and (f) for $SiO_2$), and steady state stages, parts (c) and (d) for $Cu$ and (g) and (h) for $SiO_2$, under an applied power flux equal to $q''=5\;W/cm^2$ . The
steady state condition is defined as the state in which the histogram of the surface temperature remains unchanged on average. In this state, the histogram bars fluctuate around their mean values, preserving the overall shape over time.

As can be seen, the span of the colorbars representing the temperature is markedly different, indicating small and large temperature gradients in the $Cu$ and $SiO_2$
cases, respectively. In addition, cold spots are observed in the $SiO_2$ case.
These cells correspond to inactive cavities during the reconstruction
period of the superheated microlayer. Because $SiO_2$ has a relatively low thermal
diffusion coefficient, adjacent cells cannot rapidly respond to the existing thermal gradient, delaying the reactivation of boiling cavities. As will be
confirmed later, a poor thermal diffusion coefficient affects the recovery of the
superheated microlayer after bubble departure and thus boiling performance as
a heat transfer mechanism.

In both heater materials, immediately after bubble departure, the associated cell is suddenly cooled as cold liquid from the bulk
quenches the heater surface. During this process, a $Cu$ cell is relatively less
affected than a $SiO_2$ cell due to its large stored energy, a result of its high heat capacity. Moreover, $Cu$ cells can rapidly recover the lost energy
from neighbour cells thanks to the material's high thermal diffusion coefficient.

Upon cooling less efficiently, the $SiO_2$ heater attains a higher mean temperature than the $Cu$ heater, as illustrated in Figs. \ref{Temp-Bubbles} (a-d) and (e-h) for $Cu$ and $SiO_2$ heaters, respectively. Furthermore, upon comparing these figures, it becomes evident that temperature fluctuations are more pronounced in the $SiO_2$ heater than in the $Cu$ heater. These variations are particularly noticeable in the active cavities, where bubbles grow and subsequently depart from the surface.

\subsection{Steady-state stage:  Macroscopic behavior}
In this section, focus is placed on certain macroscopic characteristics of the boiling process once the systems have reached steady state. Moreover, the effect of the applied power is discussed.

\begin{figure}[htp]
\centering
\includegraphics[width=16pc]{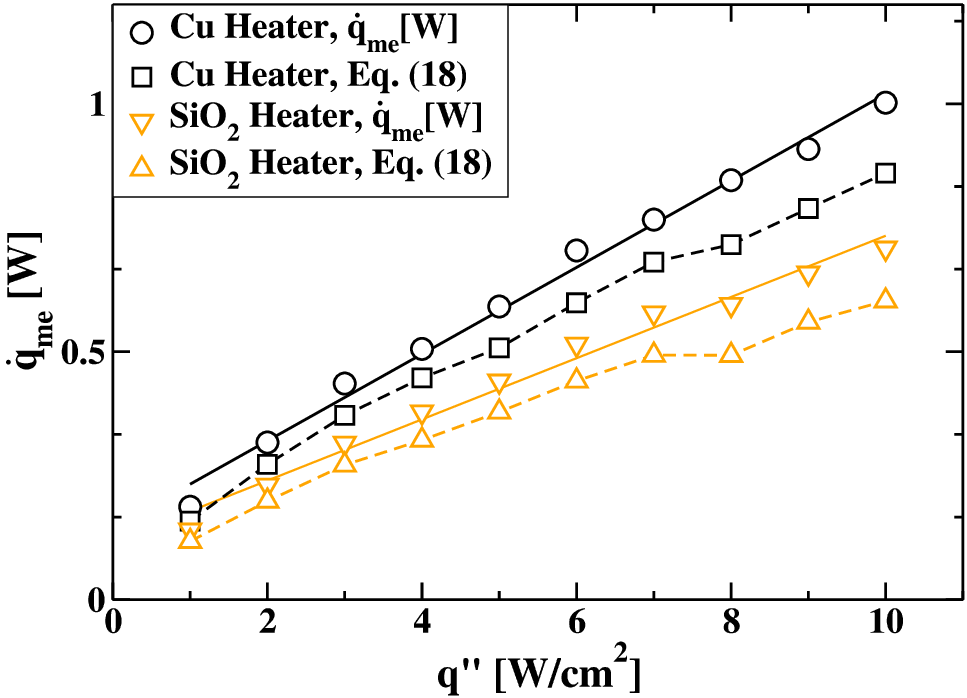}  
\caption{\label{frecxActsites} Total microevaporation heat transfer ($\dot{q}_{me}$) vs. input power flux ($q''$) for $Cu$ (black circles) and $SiO_2$ (orange upside-down triangles) heaters. Linear regressions are fitted to the data, revealing linear trends, consistent with the behavior reported in \cite{Jakob41}.}
\end{figure}

In his experiments Jakob \cite{Jakob41} observed a linear relationship between the input power flux and the power extracted by means of the microevaporation process ($\dot{q}_{me}$) within a power flux range of $0.95$ to $5.8\; W/cm^2$. Illustrating this behavior, Fig. \ref{frecxActsites} depicts $\dot{q}_{me}$ as a function of the input power flux ($q''$) for both $Cu$ and $SiO_2$ calculated with the model. Results are averaged over ten heaters with equivalent Gaussian cavity-size distributions.
As input power increases, the heat extracted via the microevaporation mechanism rises for both types of heaters. Remarkably, for a given input power value, the cooling through this mechanism is higher for $Cu$ heaters compared to $SiO_2$ heaters. The amount of power extracted through this mechanism amounts to 10 and 7$\%$ of the input total power, respectively.
The $Cu$ heater, with higher thermal conductivity and diffusivity, exhibits enhanced cooling efficiency through this mechanism compared to the $SiO_2$ heater having lower thermal conductivity and diffusivity values.
The contribution of the microevaporation cooling mechanism $\dot{q}_{me}$ (See Eq. \ref{q_me}) depends on both the mean frequency of the bubbles, denoted as $\nu$, and the mean number of active sites, denoted as $n'$. Since the detachment radii of the bubbles remain the same $\dot{q}_{me}$ can be estimated by equation \ref{dotq_me}, see  \cite{VanStralen75}.

\begin{equation}\label{dotq_me}
 \dot{q}_{me} = \rho_v h_{lv} \frac{\pi \left(2r_d \right)^3}{6} \nu n' \mbox{.}
\end{equation}

Figure~\ref{frecxActsites} presents Eq.~(\ref{dotq_me}) plotted against $q''$ for the $Cu$ and $SiO_2$ heaters, with the model computing $\nu$ and $n'$ by averaging over time windows corresponding to the steady-state condition. Comparing these results to $\dot{q}_{me}$, the present model exhibits good agreement with Eq. (\ref{dotq_me}). The product of $\nu$ and $n'$ is greater for $Cu$ than for $SiO_2$ heaters, therefore confirming materials with greater thermal conductivity and diffusivity achieve better refrigeration through nucleate boiling due to the enhanced mechanism of microevaporation.

Finally, for saturated pool boiling, a recommended empirical correlation for the product of bubble frequency and departure diameter $D_d=2r_d$ is given by the following equation \cite{VanStralen75}:

\begin{equation}\label{eq18}
\nu D_d = 1.18 \frac{t_g}{t_g + t_d}\left[\frac{\sigma g ( \rho _l - \rho _v ) }{\rho _l^2}\right]^{1/4}  \;\mbox{.}
\end{equation}

Here, $t_g$ represents the average growth time of a bubble, and $t_d$ is the average time lag between consecutive bubbles. It is important to note that this latter time can be greater than the waiting time $t_w$ defined below Eq. (\ref{twmin}), as the microlayer may recover, and the cavity is influenced by the temperature of neighboring cavities. The values of $t_g$ and $t_d$ can be determined by calculating the evolution of the bubble radius at each active nucleation site, as predicted by the model. 

\begin{figure}[htp]
\centering
\includegraphics[width=16pc]{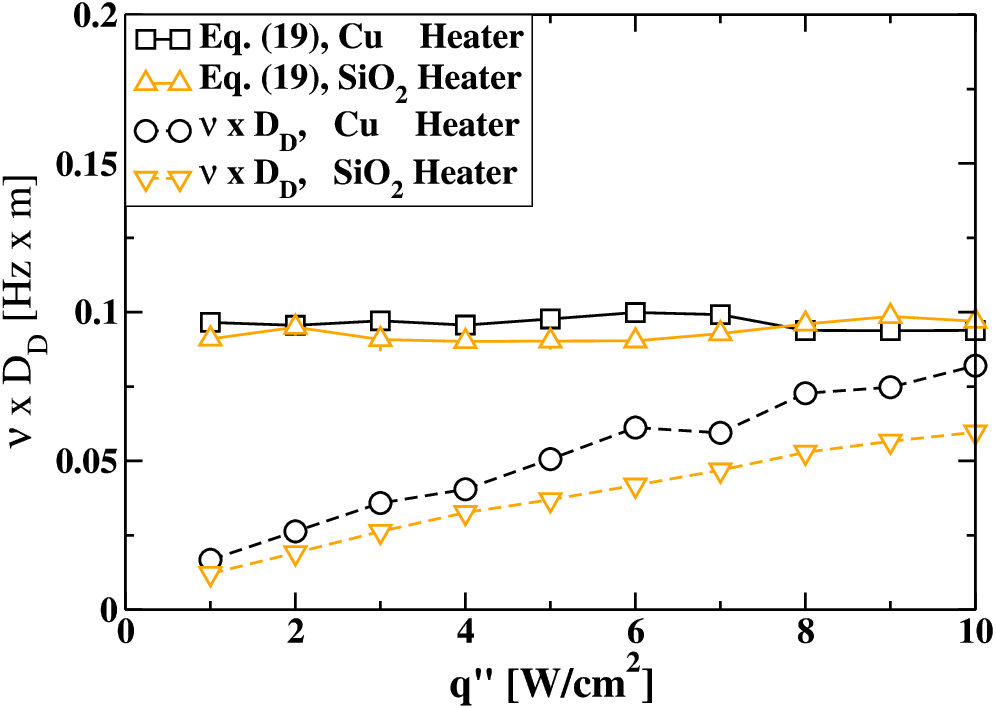}  
\caption{\label{eq} Product of the bubble mean frequency times the departure size as a function of $q''$, for both $Cu$ and $SiO_2$ heaters. Square and up triangle symbols correspond to the values obtained using
Eq. (\ref{eq18}), with  $t_g$ and $t_d$ calculated by simulations, for $Cu$ and $SiO_2$ heater, respectively. The parameters are consistent with those in Fig. \ref{frecxActsites}}
\end{figure}

In Fig. \ref{eq}, Eq. (\ref{eq18}) is plotted against $q''$ for $Cu$ and $SiO_2$ heaters. Additionally, $\nu D_d$ as a function of $q''$ is shown in the same figure for comparison with Eq. (\ref{eq18}). The results show a better fit as the input power increases. In fact, Eq. (\ref{eq18}) serves as an approximation to estimate the order of the involved magnitudes.

\begin{figure}[htp]
\begin{tabular}{cc}
\centering
\includegraphics[width=12.6pc]{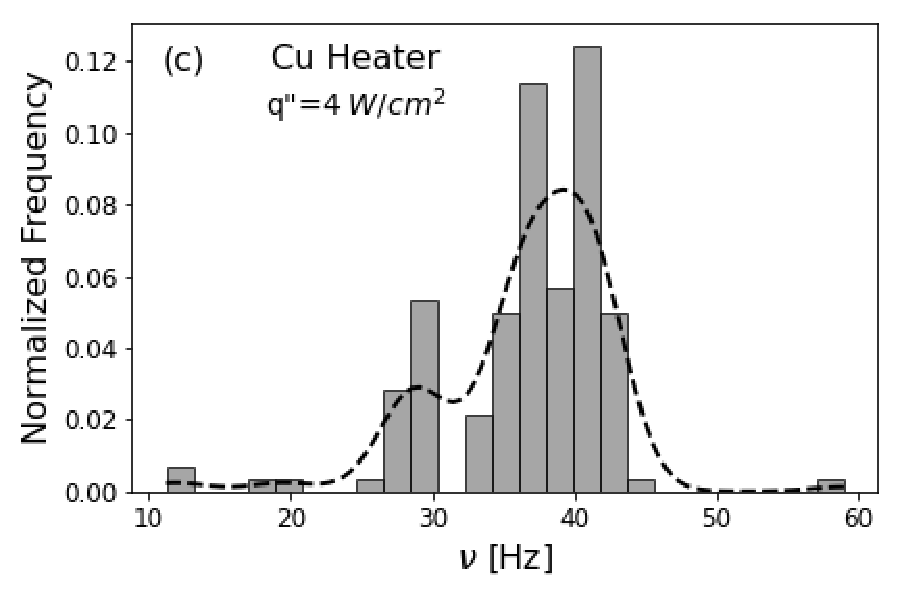}  &
\includegraphics[width=12.6pc]{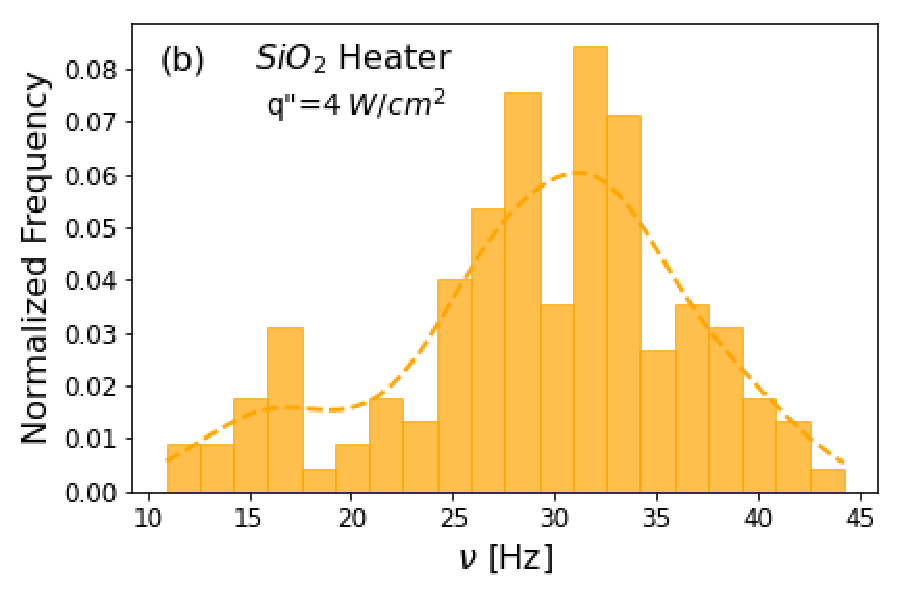}  \\
\includegraphics[width=12.6pc]{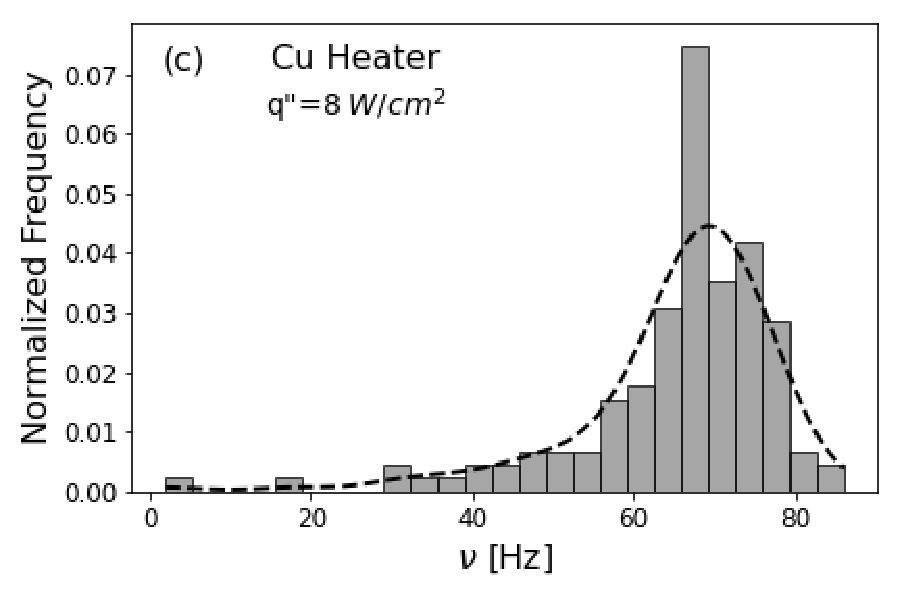}  &
\includegraphics[width=12.6pc]{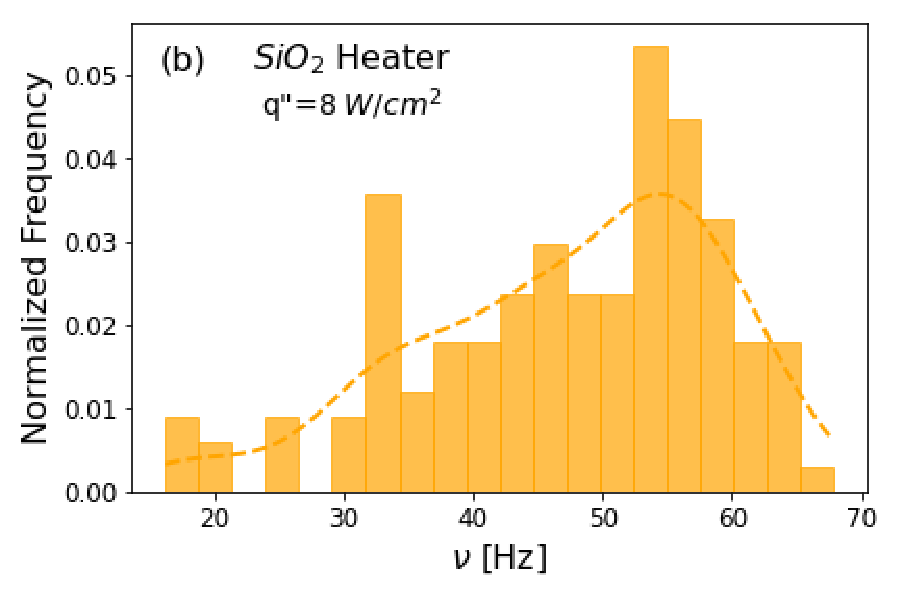}  \\
\end{tabular}
\caption{\label{histo} Histograms of frequencies: (a) and (c) for $Cu$, and (b) and (d) for $SiO_2$ heaters at two different input powers ($q''=4$ and, $8\;W/cm^2$). Solid lines correspond to kernel density estimations (KDE) of the histograms, revealing two peaks in each histogram.
}  
\end{figure}

Figure \ref{histo} presents the statistical distribution of observed boiling frequencies for $Cu$ and $SiO_2$ heaters under two different power fluxes ($q''=4$ and $8\;W/cm^2$). For each case, ten simulations were performed for both $Cu$ and $SiO_2$ heaters. The solid lines correspond to the kernel density estimations (KDE) of the histograms.

As observed, distinct patterns emerge, indicating that the boiling structures exhibit different topologies depending on the heater material. In particular, the KDE for the $Cu$ heater shows two well-defined peaks, whereas the $SiO_2$ heater displays a broader, more diffuse distribution at lower power flux.

\subsection{Steady-state stage:  Microscopic behavior}

\begin{figure}[htp]
\centering
\includegraphics[width=16pc]{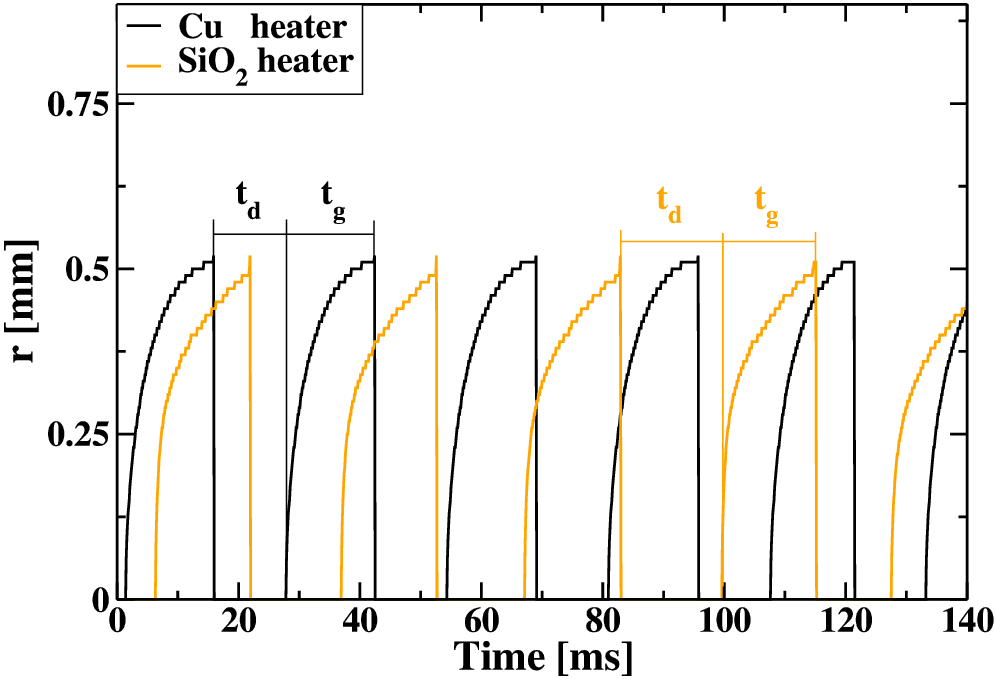}
\caption{\label{radii2} Bubble radius $r(t)$ vs. time in the steady state with adjusted time zero, for both Cu and $SiO_2$ heaters. The parameters are consistent with those in Fig. \ref{frecxActsites}. Quiescent states are observed in bubbles growing on the $SiO_2$ heater (orange line). $t_d$ and $t_g$ are the growth time of the bubble and the lag time between consecutive bubbles.}
\end{figure}

\begin{figure}[htp]
\begin{tabular}{cccc}
\centering
\includegraphics[width=12.6pc]{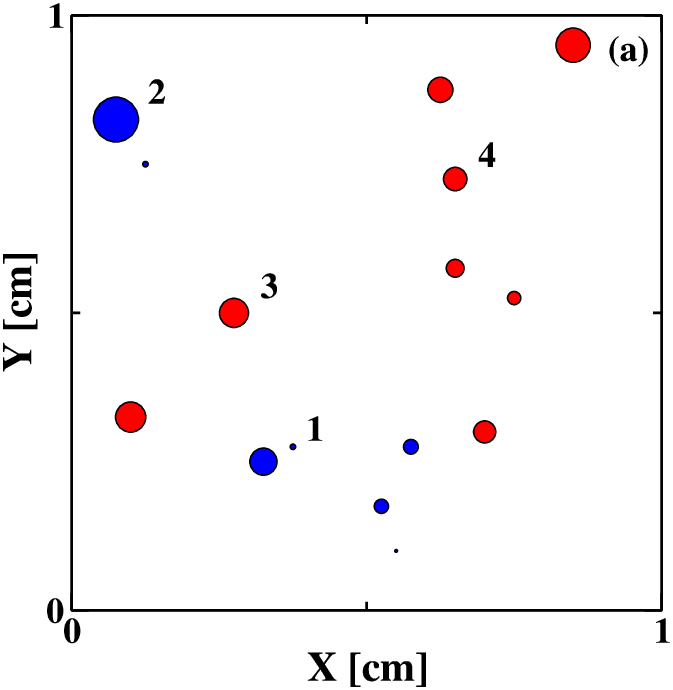}  &
\includegraphics[width=13.2pc]{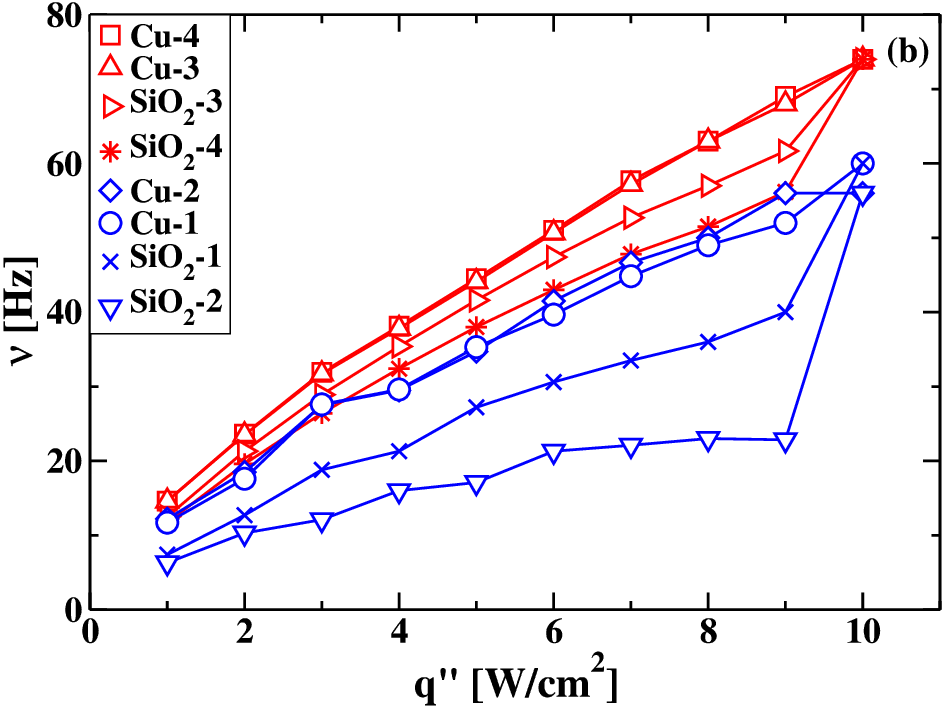}  \\
\end{tabular}
\begin{tabular}{cc}
\centering
\includegraphics[width=12.6pc]{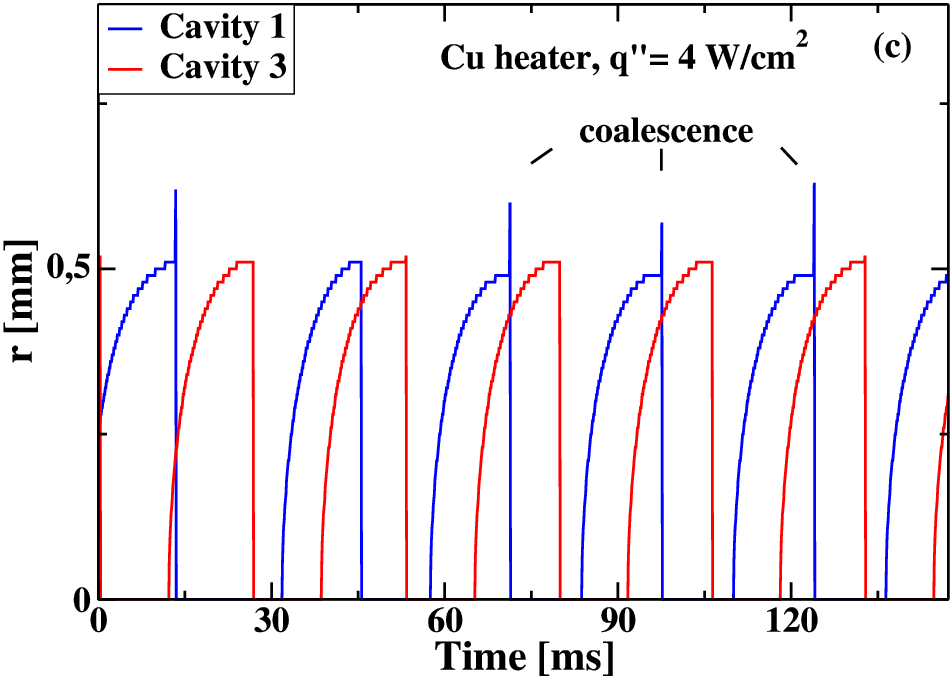}  &
\includegraphics[width=12.6pc]{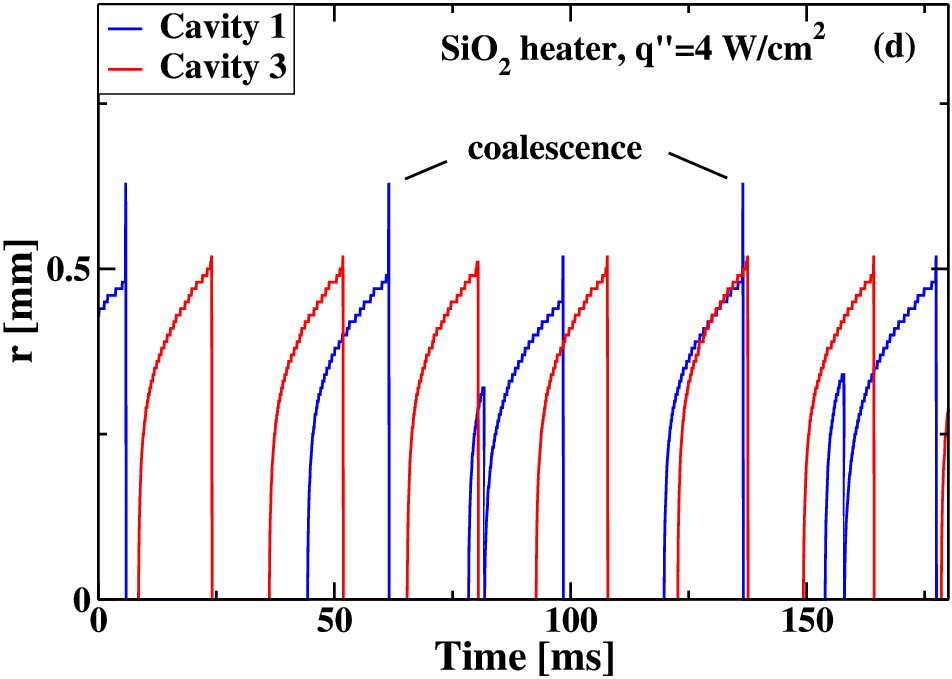}  \\
\end{tabular}
\caption{\label{regimes} (a) Active cavity positions and their proportionally amplified radii on $Cu$ and $SiO_2$ heaters.
(b) Steady bubble frequency $\nu$ as a function of $q''$ for four cavities on $Cu$ and $SiO_2$ heaters with the topology shown in part (a).
(c)-(d) Bubble radius growth $r(t)$ for cavities 1 and 3 in the steady-state regime at $q'' = 5\;W/cm^2$: (c) $Cu$ heater, (d) $SiO_2$ heater. In all cases, blue curves and circles correspond to active cavities in the low-frequency regime, while red curves and circles correspond to active cavities in the high-frequency regime.}
\end{figure}

To elucidate the microscopic behavior of the boiling process for the two different heater materials, the evolution of the bubble radius $r(t)$ from a given cavity is compared for both $Cu$ and $SiO_2$ heaters at an applied power flux of $4\; W/cm^2$. Here, $t_g$ and $t_d$ denote the bubble growth time and the time lag between consecutive bubbles, respectively.

When comparing the bubble growth evolution in the same cavity for the two materials, clear differences are observed. Both $t_g$ and $t_d$ are larger for the $SiO_2$ heater than for the $Cu$ heater. This is explained by the lower thermal diffusivity of $SiO_2$, which delays the reconstruction of the superheated microlayer and, consequently, the reactivation of the cavity, leading to a longer $t_d$.

In addition, during the bubble growth period, quiescent states are observed for bubbles on the $SiO_2$ heater. These occur when the cavity is temporarily inactive while still retaining an attached vapor bubble. Such a situation arises when bubble growth is halted due to a collective effect, where neighboring cells cannot supply sufficient heat to sustain vapor production. This phenomenon is directly related to the low thermal conductivity and diffusivity of $SiO_2$, resulting in distinct $r(t)$ profiles for bubbles on $Cu$ and $SiO_2$ heaters (see Fig.~\ref{radii2}).

A collective effect between neighboring cavities can be identified when modeling topologically identical $Cu$ and $SiO_2$ heaters operated at different heat flux values ($q''$). Figure~\ref{regimes}(a) shows the location of the cavities on each heater, where the size of the discs is proportional to the amplified radii of the cavities. Blue discs represent cavities operating at low frequency, while red discs correspond to high-frequency operation. Indeed, the bubbling frequency depends strongly on heater topology. Figure~\ref{regimes}(b) shows the bubble frequency $\nu$ for the four cavities labeled 1 to 4 in part (a) for both heater materials. Two distinct regimes, low and high frequency, are identified for each heater. The low-frequency regime is attributed to a collective phenomenon among closely spaced active cavities.

The temporal evolution of the bubble radius $r(t)$ for cavities 1 and 3 at $q'' = 4\;W/cm^2$ is shown in Fig.\ref{regimes}(c) for the $Cu$ heater and Fig.\ref{regimes}(d) for the $SiO_2$ heater. In both cases, the blue line corresponds to active cavity 1 in the low-frequency regime, while the red line corresponds to active cavity 3 in the high-frequency regime. Bubbles generated in cavity 1 sometimes coalesce with bubbles from neighboring cavities, and their growth in the low-frequency regime is irregular for both heaters. In contrast, bubbles from cavity 3 exhibit higher frequency and greater regularity. Differences in the growth profiles between heater materials are also evident, consistent with the results shown in Fig.~\ref{radii2}.

\section{Conclusions}

We analyze the effect of the intrinsic properties of the heater material on
nucleate boiling through a model introduced in this work. This model incorporates a number of expressions suggested in the literature, aimed to describe the
various mechanisms involved in these phenomena. In experiments, it is impossible to change one parameter without modifying the others. In other words, even
reproducing these experiments under the same conditions poses a challenge. For
instance, the vapor trapped in surface imperfections changes significantly from
one trial to another, making it difficult to study which mechanism is predominant when comparing results. Numerical simulations allow us to overcome this
problem and study changes by modifying one parameter at a time.

In this study, we compare two heaters made from different materials: $Cu$ , a
good thermal conductor, and $SiO_2$ , a poor conductor. Despite having identical
morphological properties, including cavity size distribution and contact angle
with the refrigerant, significant differences in nucleate boiling performance are
observed. The most notable differences include:

\begin{itemize}
 \item The $SiO_2$ heater exhibits a longer transient stage with higher mean temperatures and greater variation compared to the Cu heater. This is attributed
to $SiO_2$ 's thermal diffusion coefficient being two orders of magnitude lower
than that of $Cu$ .

\item In the steady state, the $SiO_2$ heater continues to show higher mean temperatures and greater
variability than the $Cu$ heater. Both heaters maintain a constant number of
active cavities, but $SiO_2$ heater has more active cavities than $Cu$ heater at
low power flux.

\item Despite the higher surface temperature in $SiO_2$ heaters, the superheated microlayer seems to be harder to rebuild after bubble detachment, leading to
lower boiling frequencies compared to $Cu$ heater at the same power flux.

\item The observed linear relationship between input power flux and power extracted via microevaporation, as noted by Jakob, is confirmed, with $Cu$
heaters showing greater cooling eficiency (10 \% power extraction through microevaporation) compared to $SiO_2$ (7 \% ). 

\item Histograms and kernel density estimations (KDE) of boiling frequency distributions reveal distinct patterns: $Cu$ heaters have two clear frequency peaks at low power flux,
while $SiO_2$ heaters show a more diffuse distribution, indicating differences in
boiling regularity and topology.  

\item Quiescent states occur during bubble growth on $SiO_2$ heaters, where bubbles
remain attached but growth halts due to insufficient power from neighboring
cells, a result of $SiO_2$'s low thermal conductivity and diffusivity.

\item When varying power flux, both heaters show two distinct bubble frequency
regimes: low and high. The low-frequency regime is attributed to collective
effects among closely located active cavities. Boiling behavior is irregular in
this regime for both heaters, while some cavities exhibit higher frequency and
regularity.

\end{itemize}

In summary, the model results demonstrate that variations in the heater’s thermal properties, particularly thermal diffusivity, affect not only the overall boiling performance but also several intrinsic characteristics of the process, including bubble frequency, boiling topology, boiling regularity, and cavity reactivation speed. A key contribution of this work is the demonstration that a low-computational-cost model can simulate the collective behavior of multiple cavities, capturing the emergence of spatial heterogeneity, where some cavities bubble at high frequencies while others remain at low frequencies or quiescent, especially at low input power flux values.

\section{References}
\bibliographystyle{unsrt}  
\bibliography{mono5b.bib}





\end{document}